# The Vocal Signature of Social Anxiety: Exploration using Hypothesis-Testing and Machine-Learning Approaches

Or Alon-Ronen[1], Yosi Shrem[2], Yossi Keshet[2], Eva Gilboa-Schechtman[1]

[1] Department of Psychology and the Gonda Multidisciplinary Brain Research Center, Bar-Ilan University, Ramat Gan, Israel

[2] Facullty of Electrical and Computer Engineering, Technion–Israel Institute of Technology, Israel

Correspondence: Or Alon-Ronen, Department of Psychology, Bar Ilan University, Ramat-Gan, 5290002, Israel; E-mail: or.alon1990@gmail.com

## Abstract

**Background** Social anxiety (SA) is a common and debilitating condition, negatively affecting life quality even at sub-diagnostic thresholds. We sought to characterize SA's acoustic signature using hypothesis-testing and machine-learning (ML) approaches.

**Methods** Participants formed spontaneous utterances responding to instructions to refuse or consent to commands of alleged peers. Vocal properties (e.g., intensity and duration) of these utterances were analyzed.

**Results** Our prediction that, as compared to low-SA (n=31), high-SA (n=32) individuals exhibit a less confident vocal speech signature, especially with respect to refusal utterances, was only partly supported by classical hypothesis-testing approach. However, the results of the ML analyses and specifically the decision tree classifier were consistent with such a speech patterns in SA. Using a Gaussian Process (GP) classifier, we were able to distinguish between high- and low-SA individuals with high (75.6%) accuracy and good (.83 AUC) separability. We also expected and found that vocal properties differentiated between refusal and consent utterances.

**Conclusions** Our findings provide further support for the usefulness of ML approach for the study of psychopathology, highlighting the utility of developing automatic techniques to create behavioral markers of SAD. Clinically, the simplicity and accessibility of these procedures may encourage people to seek professional help.

# 1. Introduction

Social Anxiety Disorder (SAD) is a prevalent condition involving marked anxiety about social or performance situations in which an individual is exposed to possible scrutiny by others. Individuals with SAD fear that in such situations they will act in a way that will be negatively evaluated by others (DSM-V, APA, 2013). SAD is associated with severe psychological, interpersonal, and professional consequences (e.g., Ruscio et al., 2008). Individuals with SAD suffer from significant functional impairment in multiple aspects of their daily lives: their academic trajectory is frequently interrupted, they are employed below potential, their social functioning suffers, and their quality of life is impaired (Aderka et al., 2012; Wong, Server & Beidel, 2012).

At the present, the diagnosis of SAD relies on self-report questionnaires and diagnostic interviews. Each of these diagnostic techniques has several limitations. Self-report questionnaires depend on the respondent's credibility, and may be affected by desirability bias (Demetriou, Ozer & Essau, 2014). Diagnostic interviews are comprehensive, but required training for these interviews is extensive, and costly. Moreover, such interviews may be daunting for individuals high in SA as they require a feared interaction with another person. Therefore, diagnosing SAD based on objective indicators may significantly enhance the accessibility of diagnosis, and improve our ability to identify individuals in need of clinical attention.

One promising way to achieve this goal is to objectively analyze vocal data -- speech. Experimental studies using vocal speech properties have shown great promise in differentiating between high-SA (HSA) and low-SA (LSA) individuals using classic hypothesis-testing approaches (e.g., Galili et al., 2013; Weeks et al., 2011; Weeks et al., 2012). In addition, the use of ML algorithms to distinguish between groups based on vocal analysis is a burgeoning and promising new field, with a list of impressive achievements. Indeed, this approach has been useful in diagnosing Parkinson and Alzheimer's diseases with impressive accuracy (96.4% and 87%, respectively; Erdogdu Sakar, Serbes & Sakar, 2017; König et al., 2015). In addition, there are multiple studies utilizing this approach for the diagnosis of emotional disorders, again with impressive success: depression with an accuracy of 79%- 87%, (Cohn et al., 2009; Espinola, Gomes, Pereira & dos Santos, 2020), PTSD with an accuracy of 75%-77% (Banerjee et al., 2019; Vergyri et al., 2015) and bipolar illness with an accuracy of 82% (Maxhuni et al., 2016). Although studies show differences between HSA and LSA participants in specific acoustic properties (e.g., Galili et al., 2013; Weeks et al., 2011; Weeks et al., 2012), only one study used ML algorithms to differentiate between these groups (Salekin, Eberle, Glenn, Teachman & Stankovic, 2018). In this study, participants (n=105) were asked to prepare and deliver a brief speech in front of a camera. The ML analysis of the resulting vocal data achieved an accuracy of 90%.

In the present study we sought to expand this important line of research using ecological, everyday speech. Specifically, we sought to examine the acoustic properties of speech in LSA and HSA individuals using two complementary approaches: Theory-driven, classical hypothesis-testing approach, and a data-driven approach. Within the *theory-driven approach* we were guided by the evolutionary model of Gilbert which views SA as a disturbance of the social status system (Gilbert, 2001). Evolutionary approaches hypothesized and found that, as compared to LSA individuals, HSA individuals evaluate themselves as low in social attractiveness, and thus fear of making bids for status or approval. HSA individuals appear to believe that claims for social status claims may lead to social exclusion or defeat, and to prevent these undesirable outcomes, HSA (vs. LSA) individuals seek to de-escalate social competition by using submissive or appeasing behaviors (Aderka, Weisman, Shahar & Gilboa-Schechtman, 2009; Schneier, Kent, Star & Hirsch, 2009; Weeks, Heimberg & Heuer, 2011). For example, as compared to LSA individuals. HSA individuals make greater use of excuses and apologies (Edelman et al., 1987), express more self-doubts during a disagreement (Oakman, Gifford & Chlebowsky, 2003) and suppress anger feelings (Erwin, Heimberg, Schineier, & Leibowitz, 2003). However, most of the evidence for negative association between SA tendencies and dominance expression is based on subjective and observational measures (e.g., Johnson, Leedom, & Muhtadie, 2012).

In contrast, *data-driven approaches* are designed to make the most accurate predictions possible for new (test) data, based on known properties learned from the training data using ML techniques. Whereas relative simplicity and good interpretability are basic features of the theory-driven approach, ML models often have weak interpretability. Combining the two approaches may allow us to better understand vocal patterns in SA without compromising on the accuracy of our predictions.

### *The present study*

The primary aim of the current study was to differentiate between HSA and LSA individuals based on the acoustic properties of their speech based on performance in motivationally directed speech. To this end, participants were presented with command sentences (e.g., "Bring me the notebook!"). uttered by a "peer." Participant's task was to <u>refuse</u> (dominant speech action) or agree (non-dominant speech action) the peer request based on visual cue (background color). Participants' speech utterances were recorded throughout the task. Our secondary aim was to classify the utterance based on their motivational intent (refusal, acceptance).

### *Speech Properties*

The speech properties in our study included seven measures. ***Fundamental Frequency (F0)*** represents the rate of vocal cords vibration during speech measured in Hz, and subjectively perceived as pitch. Pitch differs widely between women and men, which is estimated to average around 220 and 130 Hz, respectively (Petersen & Barney, 1952). Robust evidence indicates that mF0, the mean pitch, is negatively correlated with self- and other-rated social and physical dominance (Jones, Feinberg, DeBruine, Little, & Vukovic, 2010; Ohala, 1984; Puts, Gaulin & Verdolini, 2006; Puts, Hodges, Cardenas & Gaulin, 2007). Weeks and colleagues found higher levels of mF0 in SAD individuals during a public speech (Weeks et al., 2012), a tendency which appears to subside following pharmacological treatment (Laukka et al., 2008). Previous studies document that participants who rated themselves as higher in dominance had lower F0-std (Leongoamez et al., 2017). Moreover, males' voices lower in F0 std were perceived to be physically dominant by other males (Hodges-Simeon, Gaulin & Puts, 2010).

*Intensity* represents the effort used by the speaker to produce speech. It is measured in DB, and it is subjectively perceived as voice loudness. Intensity is positively associated with self- and other-rated social dominance in the production of spontaneous speech (Leinonen, Hiltunen, Linnankoski & Laakso, 1996; Linnankoski, Leinonen, Vihla, Laakso, & Carlson, 2005; Tusing & Dillard, 2000). Voices characterized by greater *variance in intensity* are perceived as more dominant (Tusing et al., 2000). Furthermore, participants' vocal parameters following a power challenge (i.e., social threat) showed a decrease in their vocal intensity. This finding is consistent with a de-escalating strategy (Bugental, Beaulieu, Schwartz & Dragosits, 2009). In addition, Galili et al. (2013) showed that high, as compared to LSA males, have lower vocal intensity. This research also documented that SA tendencies are associated with decreased vocal intensity in request utterances, both in males and in females, and with less increase of vocal intensity in command utterances in males.

*Jitter and Shimmer* measure (respectively) the cycle-to-cycle variations in F0 and intensity (Ferrand, 2001). Having higher shimmer and jitter has been correlated with increased pathogenesis of muscles related to voice (Shao, MacCallum, Zhang, Sprecher & Jiang, 2010), which means that increased levels of jitter and shimmer may be signs of weakness. Therefore, individuals with lower jitter and shimmer may sound stronger and more dominant. SAD participants were found to have higher jitter and shimmer when answering questions, compared to control participants. (Silber-Varod, Kreiner, Lovett, Levi-Belz & Amir, 2016).

***Speech Duration*** refers to the duration (in seconds) of the uttered speech**.** When SAD and control participants were asked to produce free responses to six identical questions. responses of individuals with SAD lasted longer than the responses of control participants, due to both more speech and more silences (Silber-Varod et al., 2016). Speaking for a longer time increased dominance ratings for those who spoke with content rated as dominant, but decreased dominance ratings for those who did not use dominant-sounding language (Hodges-Simeon et al., 2010).

***Speech onset (promp-to-start)*** refers to the time that passes from the beginning of the recording to the moment that the participant starts to speak. HSA individual employ a range of strategies to prevent feared social consequences of appearing silly or unprepared. One of these strategies is to mentally rehearse sentences before speaking (Cuming et al., 2009). For this reason, speech onset may be delayed in HSA vs. LSA individuals.

***Relative Silence*** refers to the total duration of unvoiced segments divided by the total duration of the speech sample. Speech-fearful and SAD participants paused more often and for longer durations than did the control group (Lewin, McNeil & Lipson, 1996). This tendency changed following treatment: participants who reported lowered SA levels demonstrated decreased relative silence post-treatment (Laukka at el., 2008).

**1.2 Hypotheses**

We tested two predictions using both classical hypothesis testing and ML algorithms approach. First, we expected HSA to be associated with lower vocal dominance (higher mF0; higher F0 std; lower mean Intensity; lower intensity std; higher jitter; higher shimmer) as compared to LSA, across utterance types (*decreased confidence in SA* hypothesis, H1). We further expected HSAs but not LSA to be characterized by longer prompt-to-start and higher proportion of relative silence. Moreover, we expected it to be easier to distinguish between HSA to LSA individuals based on refusal, as compared to the consent utterances, due to the difficulty of HSA individuals with expression of dominance.

An additional purpose of the current study was to distinguish between refusal and consent utterances, which are associated with dominant and submissive intents, respectively (distinct motivational profiles, H2). Spoken utterance classification which involves robustly classifying the intent of a person is an investigated topic in computer science, driven by the motivation to improve the communication between humans and machines (e.g., Price, Mehrabani & Bangalore, 2020).

Based on previous findings, we expected consent utterances to be associated with lower vocal dominance (higher F0 std; lower mean Intensity; lower intensity std; higher jitter; higher shimmer) as compared to refusal utterances (Gallili et al., 2013; Leinonen et al, 1996; Linnankoski et al., 2005). The demonstration of high vocal intensity levels when uttering commands was reported in previous studies (Galili et al., 2013; Leinonen et al, 1996; Linnankoski et al., 2005) and has been linked to the expression of strong negative feelings or dominance attempts (e.g., Kimble & Musgrove, 1988), whereas lower vocal intensity, found in uttering request, and was perceived as signaling lower social status (Hall, Coats & Lebeau, 2005). Although lower vocal dominance is also characterized by higher mF0, we did not formulate a specific assumption regarding this parameter because mF0 level was found to increase in the expression of a variety of emotional states and motivational intents (Banse & Scherer, 1996; Galili et al., 2013).

## 2. Method

### 2.1 Participants

Participants (N=118, 69 women) were university students who took part in the study in exchange for 60 NIS (approximately $13) or academic credit. Participants were recruited through the Psychology Department Subject Pool, as well as from advertisements in billboards on campus and electronic forums. Two participants who did not meet the inclusion criteria of fluent Hebrew speech were excluded from the study. Participants ranged in age from 20 to 42, with a mean age of 23.49 years (SD=2.90). Participants' level of education ranged from 12 to 20 years, with a mean of 12.97 (SD=1.45).

### 2.2 Procedure

Participants were presented with 24 command sentences (e.g., "Bring me the notebook!", "Clean the room!"). Each sentence was followed by a picture of ostensible "peer" presented on either a blue or red background color (in total, 12 pictures had a blue background and 12 had a red background). To make the task more ecological, the participants were asked to imagine that the "peer" commanded them to do an action. The participants' task was to <u>refuse</u> the peer when the background was red and to <u>agree</u> when the background was blue (using freely created sentences). The participants' voice was recorded throughout the task. To make sure that the participants understood the instructions, the task started with several training trials, during which the experimenter made sure that the task was progressing smoothly. All the participants without exception mastered the task after completing the training trails. Following the completion of the vocal task, the participants filled out questionnaires assessing the severity of SA-severity.

*Self-Report Measures*

**Liebowitz Social Anxiety Scale (LSAS; Liebowitz, 1987).** A 24-item self-report questionnaire measuring anxiety and avoidance in social or performance situations on a 0–3 scale. LSAS has been shown to have high internal consistency, strong convergent and discriminate validity, and high test-retest reliability (Baker, Heinrichs, Kim, & Hofmann, 2002; Fresco et al., 2001). The internal consistency in our sample was 0.94.

**2.3 Recordings and Acoustic Analyses**

Recording sessions were performed individually in a quiet room. The experimenter familiarized the participants with the equipment and remained present in the room during the entire recording session. Participants' speech signals were recorded using a Sennheiser PC20 headset microphone (High Wycombe, United Kingdom). The microphone was positioned approximately 5 cm from the corner of the participant's mouth and connected directly to a desktop computer. Speech samples were recorded using the GoldWave program (Version 5.12, GoldWave, Inc., 2005), with a sampling rate set at 48 kHz (16 bit), mono channel (see Rochman and Amir, 2013 for a brief introductory tutorial on basic procedures for recording speech/voice and acquiring relevant acoustic measures).

The recordings were then analyzed using free standard audio-based tools and libraries. First, the audio recordings were truncated using Voice Activity Detector (VAD; Lee & Hasegawa-Johnson, 2007) to remove non-speaking parts. Next, using Praat and PraatIO (Boersma & Weenik, 2009; Mahrt, 2016) we extracted 18 speech properties (min F0, max F0, mean F0, std F0, intensity min, intensity max, intensity mean, intensity std, jitter, shimmer, silence, number of silences, duration, and prompt to start).

## 3. Results

### 3.1 Participant Characteristics

Participants were divided into high- and low-SA groups based on their LSAS scores. The SA severity score in the low-SA group (LSA, n=31; 19 women) were below 30 (Mennin et al., 2002), and ranged from 4 to 30, with a mean of 19.74 (SD = 7.23). The LSAS score in the high-SA (HSA, n=32; 19 women) group were above 50 (for similar clinical cutoff scores, see Lazarov et al., 2018), and ranged from 50 to 107, with a mean of 67.13 (SD = 15.17). Such grouping and severity patterns are consistent with previous research (e.g., Salekin et al., 2018), Participants with LSAS score of 30-50 were excluded from this analysis. As expected, the groups differed in LSAS scores ($F(1, 61)$= 165.03, $p < .001$) but did not differ on any other demographic variable (all $Fs<2.3$).

### 3.2 Vocal Parameters

For each vocal parameter, outliers of more than three standard deviations above or below the mean were excluded from the analysis (as in Weeks et al., 2011). Means and standard deviation of vocal parameters for men and women separately are presented in Table 1.

**Table 1**

*Means and Standard Deviation of Vocal Parameters*

|  | Men (N=47) | Women (N=69) |
|---|---|---|
| Min F0 | 87.67 (12.74) | 131.22 (23.92) |
| Max F0 | 216.15 (43.34) | 320.82 (36.66) |
| Mean F0 | 120.17 (16.10) | 195.97 (23.64) |
| Std F0 | 28.05 (10.71) | 39.21 (8.40) |
| Intensity min | 40.39 (2.32) | 37.54 (2.16) |
| Intensity max | 62.84 (5.62) | 66.35 (6.23) |
| Intensity mean | 53.61 (4.42) | 54.53 (5.04) |
| Intensity std | 5.37 (0.99) | 6.56 (1.18) |
| Jitter | 0.01 (0.003) | 0.02 (0.003) |
| Shimmer % | 13 (2) | 11 (2) |
| Jitter voice breaks | 4.66 (1.10) | 4.92 (1.23) |
| Silence msec 50 | 0.64 (0.38) | 0.66 (0.48) |
| Silence msec 100 | 0.38 (0.25) | 0.42 (0.31) |
| Silence msec 150 | 0.23 (0.17) | 0.27 (0.24) |
| Silence msec 200 | 0.14 (0.13) | 0.15 (0.18) |
| Prompt-to-start | 1.28 (0.37) | 1.5 (1.22) |
| Relative-silences | 0.06 (0.03) | 0.06 (0.03) |
| Duration | 1.38 (0.29) | 1.57 (0.49) |

### 3.3 Decreased Confidence in SA

*Top-Down (Hypotheses Based) Analyses*

Separate ANOVAs analyses were conducted for each theoretically relevant parameter: mean F0, F0 std, mean-intensity, intensity-std, jitter, shimmer, relative-silences and prompt-to-start; with SA (2 levels: LSA and HSA) as between subject variable. Because F0 differs widely between genders, we conducted this specific analysis separately for males and females. Means and standard deviation for each ANOVA analyses are presented in Table 2. Consistent with our hypothesis, HSA participants demonstrated significantly lower mean intensity than did LSA participants, (M = 52.34, SD = 4.57; M =54.82, SD = 4.39 respectively; $F(1, 61) = 3.83$, $\eta^2=.05$, $p < .05$ ). No significant SA differences were observed on any of the other vocal parameters, (all Fs<2.10, p>.069).

**Table 2**

*Means and Standard Deviation of Acoustic Parameters for High and Low SA groups.*

|  | LSA | HSA | F values | $\eta^2$ values | P values |
|---|---|---|---|---|---|
| Acoustic features |  |  |  |  |  |
| Mean F0- females | 189.62(22.36) | 200.43(25.69) | 2.04 | 0.05 | 0.08 |
| Mean F0- males | 123.02(18.30) | 117.12 (14.61) | 1.14 | 0.04 | 0.29 |
| F0 std | 36.78(11.45) | 38.12 (11.26) | 0.21 | 0.003 | 0.32 |
| Mean intensity | 54.82 (4.39) | 52.34 (4.57) | 3.83* | 0.05 | 0.02 |
| Intensity std | 6.35 (1.15) | 5.90 (1.37) | 2.02 | 0.03 | 0.08 |
| Jitter | 0.02 (0.003) | 0.02 (0.004) | 0.02 | 0.00 | 0.88 |
| Shimmer | 0.11 (0.02) | 0.12 (0.03) | 2.09 | 0.03 | 0.07 |
| Relative silence | 0.05 (0.02) | 0.06 (0.04) | 0.29 | 0.004 | 0.29 |
| Prompt to start | 1.39 (0.60) | 1.43 (0.53) | 0.34 | 0.005 | 0.27 |

*p<.05, **p<.01

We also conducted nine 2 (SA: LSA vs HSA) X 2 (sentence type: refusal vs consent, within subject) ANOVAs analyses (one for each parameter that we had a specific hypothesis on - mean F0, F0 std, mean Intensity, intensity std, jitter, shimmer, relative silence and prompt to start) to examine the pattern of differences between HSA and LSA individuals based on the refusal as compared to consent utterances. No significant SA main effects or SAX sentence type interactions were observed on any of the parameters (all Fs<3.283, p>.079).

*Data-Driven Analyses*

In the following sections, we describe the binary HSA/LSA classification process and provide a brief explanation of the ML algorithms found to differentiate between the SA groups in the most effective manner. Next, given the pronounced gender differences on vocal parameters, we present separate analyses for males and for females.

The features for each recording had been normalized on gender basis prior to feeding them into the learning process. To support new and unannotated recordings, we first classified the speaker's gender and normalize accordingly. The gender classification achieved 99% accuracy for predicting whether the speaker is a male or a female based on the raw acoustic values. Based on the gender classifier, the features were normalized and forwarded to the SA classifier.

We have examined several ML binary classifications and evaluated their performance using the following metrics: (a) *accuracy*: the ratio between the correctly predicted observations to the total observations. The question that this metric address is: from all our predictions, how many were correct? (b) *Precision*: the ratio between the correctly predicted positive observations to the total predicted positive observations. The question that this metric answer is: from all the participants that were labeled as HSA, how many were, in fact HSA? (c) *Recall*: the ratio between correctly predicted positive observations to all observations that were actually true. The question Recall answers is: from all the participants who are truly HSA, how many did we label as such? We practiced 10-folds cross-validation, whereas 10% was left for testing and the other 90% for training. The mean performance and the standard deviation of the ten runs are reported in Table 3.

**Table 3**

*LSA and HSA ML classification*

| Algorithm | Acc. | Prec. | Rec. |
|---|---|---|---|
| Decision Tree | 64.8 ± 2.1 | 0.61 ± 0.02 | 0.57 ± 0.03 |
| Nearest Neighbors (3) | 73.7 ± 2.7 | 0.73 ± 0.05 | 0.63 ± 0.05 |
| Gaussian Process | **75.6 ± 1.6** | **0.74 ± 0.03** | **0.69 ± 0.02** |
| Deep Neural Network | 68.9 ± 1.2 | 0.66 ± 0.03 | 0.61 ± 0.03 |
| XGBoost | 71.3 ± 1.5 | 0.69 ± 0.03 | 0.63 ± 0.03 |
| Logistic Regression | 61.2 ± 1.6 | 0.59 ± 0.05 | 0.39 ± 0.04 |

These results suggest that our proposed model incorporating the Gaussian Process (GP) classifier outperforms other methods, achieving 75.6% accuracy without compromising the balance between precision and recall. Hence, the model performs comparably well in detecting both HSA and LSA. To further evaluate our model's robustness, we have plotted the Receiver Operating Characteristics (ROC) curve and report the Area Under the Curve (AUC) in Figure 1. The ROC curve is created by plotting the true positive rate (also known as recall) against the false positive rate (the ratio between incorrectly predicted positive observations to all observations that were actually false) at various threshold settings. The AUC metric ranges from 0 to 1. The diagonal dotted line corresponds to AUC of 0.5 and equals to a coin-flipping prediction. All the points on the diagonal dotted line correspond to a situation where true positive rate is equal to false positive rate. All points above this line correspond to the situation where the true positive rate is greater than the false positive rate, and when AUC=1 the classifier is able to perfectly distinguish between the positive (HSA distribution) and the negative (LSA distribution) class points correctly. In other words, the higher the AUC is the better its ability to distinguish between HSA and LSA distributions. As plotted in Figure 1, the mean AUC of the 10-fold cross-validation experiments in our study is 0.83.

**Figure 1**

*AUC-ROC curve*

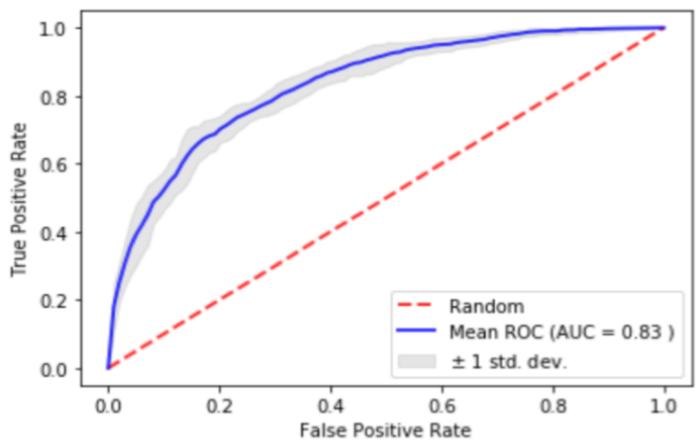

Next, we represented the model with 18 selected features that would capture both the temporal aspects of the speaker and the vocal parameters exhibited by HSA and LSA. Taking all the features into account may force our model into complex and uninterpretable regions. Moreover, while various features could be highly informative, others might be disrupting. Hence, we wanted to reduce recording representation without risking performance.

Because certain features are highly correlated (e.g., the mean and the max values of the F0) our initial assumption was to keep only a single feature relating to the same characteristic. However, filtering features as proposed has not been proven to perform better than keeping all the features. For this reason, we used another dimensionality reduction technique. This technique was based on the ANOVA F-value between a feature value and the existence of HSA, a higher score indicating a greater correlation. Next, for each number of features selected according to their importance rank, we performed the learning pipeline stated above - split the data based on gender, and then train a separate model for males and females. The mean accuracy classifying HSA/LSA of the 10-fold cross-validation procedure and the features rank is reported in Figure 2. For example, where the x-axis, the number of features, equals 4, we filter our data and keep only the four features with the highest score. The y-axis is the mean performance, e.g., the male classifier achieved an average of 66% accuracy predicting HSA and LSA with four features.

**Figure 2**

*Mean Accuracy of 10-Fold Cross-Validation Procedure and Features Rank for male and female*

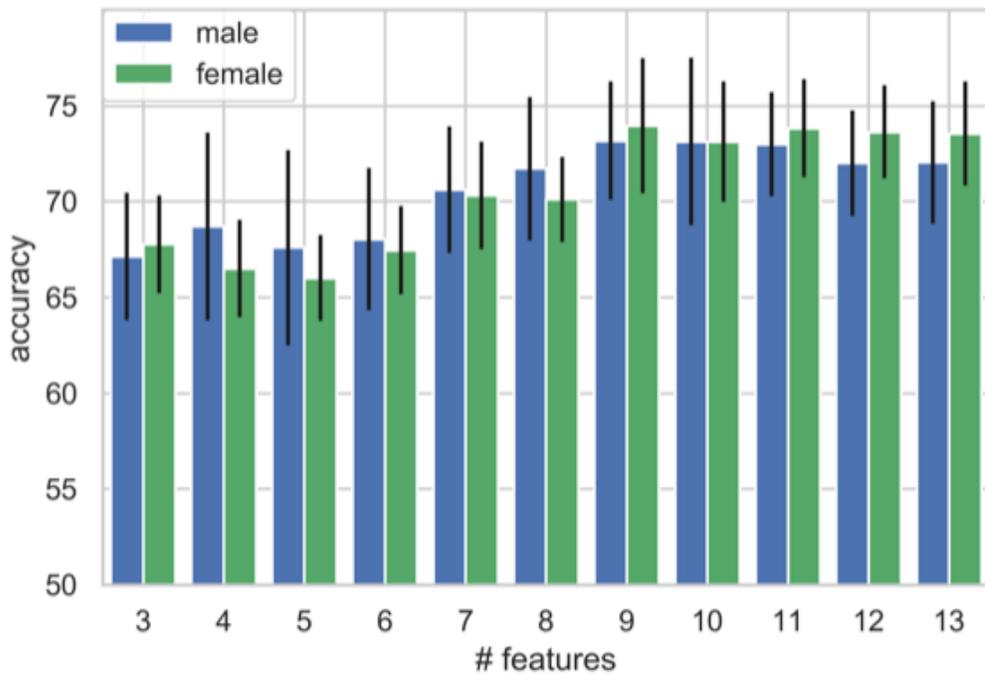

As Figure 2 clearly demonstrates, adding features contributes to the performance across genders up to a certain point (which appears to be 9 features) and more features lead to performance degradation. Best accuracy for both genders achieved using the following nine parameters (ordered by ranking): intensity mean, intensity max, min F0, intensity min, jitter voice brakes, intensity std, mean F0, duration and prompt to start.

We also explored whether HSA male speakers' vocal patterns was comparable to HSA female speakers' vocal patterns. For that purpose, we repeated the previous analysis separately for each gender. As can be seen in Figure 3, regardless of the features number, the performance achieved by gender-specific classifier outperformed the gender-unifying configuration (sharing the same features for males and females, as shown in Figure 2).

**Figure 3**

*The Mean Accuracy of the 10-Fold Cross-Validation Procedure and the Features Rank, using Gender-Specific configuration*

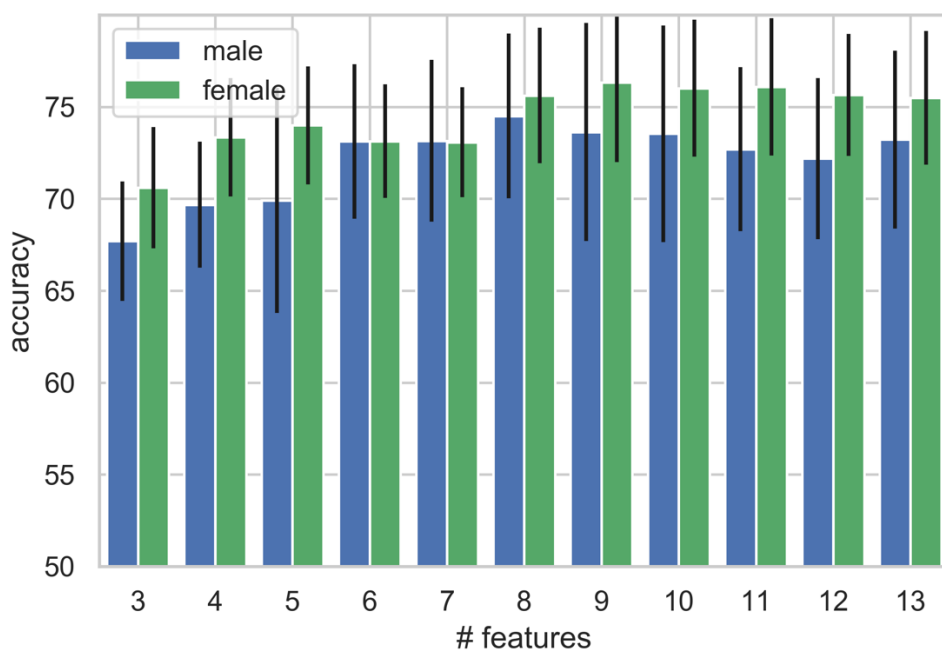

Unlike the previous configuration, the features' importance now was slightly different across genders. the female classifier utilizes the following nine best features performed (ordered by ranking): mean F0, min F0, intensity mean, prompt to start, intensity max, intensity min, jitter voice brakes, shimmer, and duration. In comparison, the male classifier presented the best results with the following eight features (ordered by ranking): mean F0, intensity max, shimmer, intensity mean, intensity std, min F0, prompt to start and jitter voice brakes.

Lastly, to visualize the differences acoustics-wise across genders diagnosed with HSA, we assessed whether a gender-specific classifier would perform equally well when providing samples of speakers from the opposite gender. In other words, we performed the same procedure as described in this section however, the results reported in Figure 4 are the accuracy achieved while training on gender A and tested on gender B. Examination of these findings suggest that the vocal patterns of LSA and HSAs exhibit gender-specific patterns. Regardless of the model's data, a suitable classifier, trained on male samples (~70%+ as in Figure 3), cannot reproduce its performance while tested on females, and vice versa.

**Figure 4**

*The Mean Accuracy of the 10-Fold Cross-Validation Procedure and the Features Rank when training on gender A and tested on gender B*

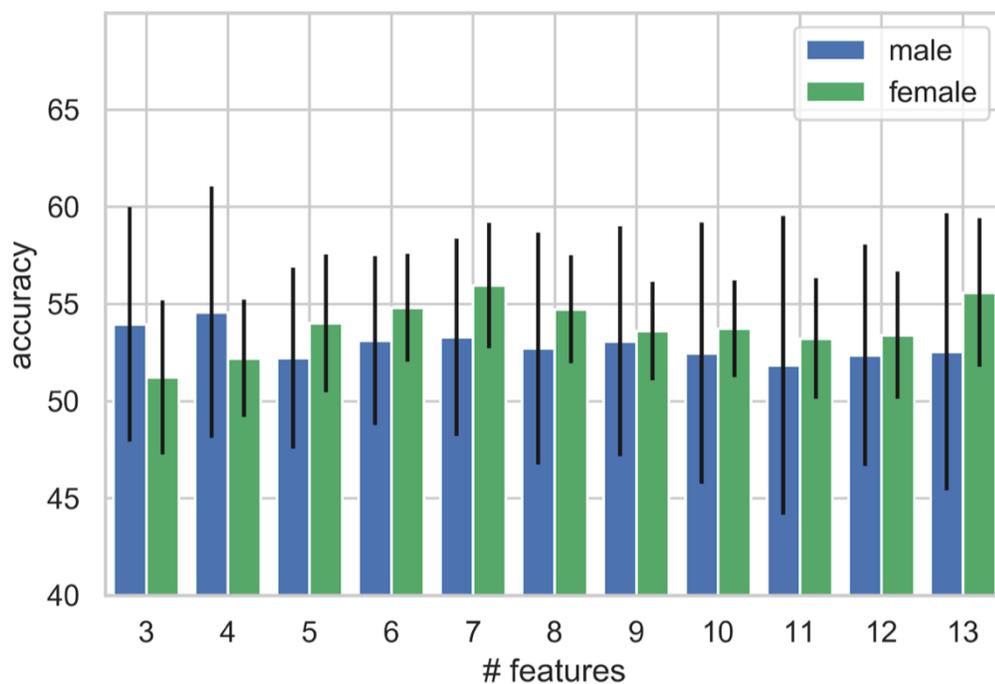

Overall, a series of analyses strongly suggests that SA affects acoustic parameters of both genders. Moreover, it appears that the acoustic signature of SA in males and females is distinct.

Generally, in ML algorithms, we are faced with an inherent tradeoff between performance and explainability of a model. Increasing the degrees of freedom or the number of parameters leads to a more complex model that is often difficult to comprehend. Thus, to emphasize that acoustics related to SA needs to be analyzed on a gender basis, we turned to probably most explainable ML algorithm - decision trees. Keeping a simplified logic harms the performance, but it is extremely helpful in gaining an initial understanding of the possible logic regarding the more complex models. The LSA vs. HSA decision tree classification is displayed in Figure 5. The decision tree makes decisions based on a sequence of questions asked about feature values. If the entropy is equal to zero it means that the sequence of questions leads to a completely homogeneous sample that all its members are either HSA or LSA participants. As can be seen from Figure 5, the decision tree created different "branches" (model) for men and women. In addition, in about 1 out of 6 of the sentences we can confidently determine whether the participants who said them were from the LSA or HSA groups (the entropy of these leaves are equal to zero). The vocal characteristics of these participants were in alignment with our hypothesis: HSA men were characterized by lower mean intensity, lower intensity variability, and higher shimmer than LSA men and HSA women were characterized by higher mean F0, lower mean intensity and longer prompt to start delay as compared to LSA women. For the other sentences, we had uncertainty (entropy>0).

**Figure 5**

*LSA vs. HSA decision tree*

Value (X=LSA, Y=HSA)

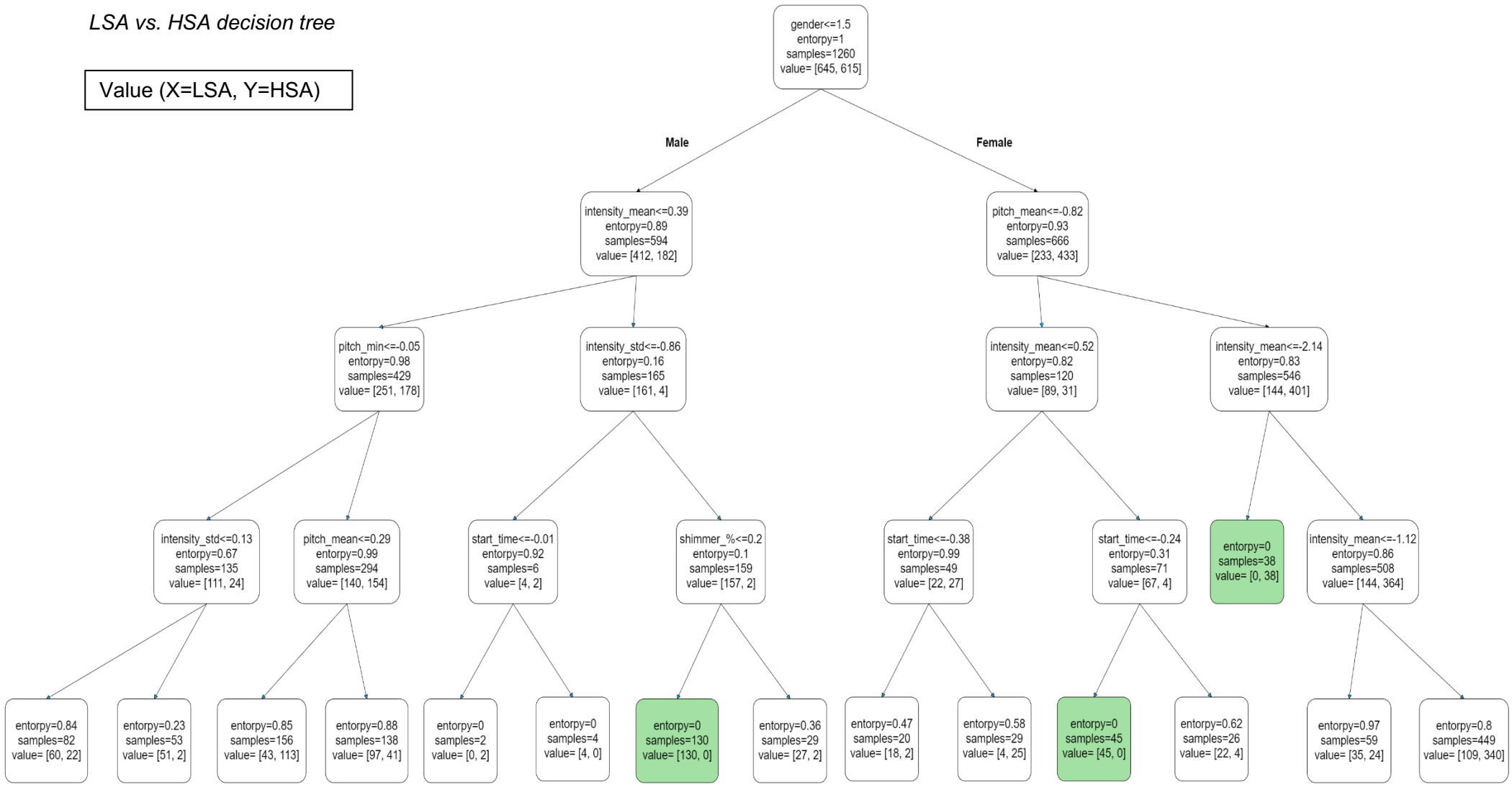

To examine whether the distinction between HSA to LSA individuals based on the refusal utterances is easier than when based on consent utterances, we divided our dataset by utterance type and sought to classify SA for each utterance type. Looking only at the recordings in which the participant uttered a consent sentence, we scored 70% accuracy. The same procedure was applied to the refusal corpus and resulted in 79% accuracy. In other words, participants with HSA were more likely to articulate differently from participants with LSA when they were asked to express refusal as compared to consent, which make it easier to differentiate between HSA and LSA using refusal utterances.

### 3.4 Distinct Profile in Refusal and Consent Utterances

*Hypotheses-Based Analyses*

We examined the difference between refusal and consent utterances in F0 std, mean Intensity, intensity std, jitter and shimmer. The pattern of findings was consistent with our predictions, such that, compared with refusal statements, consent statements were characterized by higher F0, lower mean intensity, lower intensity std, higher jitter, and higher shimmer. Means, standard deviations, and statistical comparisons are presented in Table 4.

**Table 4**

*Means and Standard Deviation of Selected Acoustic Parameters of Refusal and Consent utterances*

|                | Refusal          | Consent          | T value | Cohen's d | P value |
|----------------|------------------|------------------|---------|-----------|---------|
| F0 std         | 34.64 (11.83) a  | 39.00 (13.01) b  | 4.65    | 0.43      | 0.00    |
| Mean intensity | 54.81 (5.17) a   | 53.06 (4.82) b   | 11.33   | 1.05      | 0.00    |
| Intensity std  | 6.35 (1.49) a    | 6.08 (1.32) b    | 3.43    | 0.32      | 0.0005  |
| Jitter | 0.015 (0.004)$_a$ | 0.016 (0.004)$_b$ | 2.50 | 0.23 | 0.007 |
| Shimmer | 0.11 (0.03)$_a$ | 0.12 (0.03)$_b$ | 5.49 | 0.51 | 0.00 |

Means with differing subscripts within rows are significantly different at p<.01

*Data-Driven Analyses*

Judging by the vocal parameters alone, we attempted to construct an utterance-type classifier, predicting whether the speaker expressed consent or refusal. We have examined several ML binary classifications and evaluated their performance using accuracy, precision, and recall metrics. We practiced 10-folds cross-validation, whereas 10% was left for testing and the other 90% for training. The mean performance and the standard deviation of the ten runs are reported in Table 5.

**Table 5**

*Consent and Rejection ML classification*

| Algorithm | Acc. | Prec. | Rec. |
|---|---|---|---|
| Decision Tree | 63.5 ± 3.9 | 0.64 ± 0.04 | 0.63 ± 0.04 |
| Nearest Neighbors (3) | 63.2 ± 2.5 | 0.64 ± 0.03 | 0.61 ± 0.04 |
| Gaussian Process | **71.7 ± 2.9** | **0.72 ± 0.03** | **0.72 ± 0.02** |
| Deep Neural Network | 71.5 ± 3.1 | 0.71 ± 0.03 | 0.72 ± 0.03 |
| XGBoost | 69.2 ± 2.6 | 0.69 ± 0.03 | 0.69 ± 0.03 |
| Logistic Regression | 71.6 ± 3.5 | 0.71 ± 0.03 | 0.72 ± 0.04 |

The result suggests that our proposed model incorporating the Gaussian Process (GP) classifier outperforms other experimented methods, achieving 71.7% accuracy without compromising on the balance between precision and recall. Hence, the model performs comparably well on both detecting consent and refusal. To further evaluate our model robustness, we have plotted ROC-AUC curve in Figure 6. As can be seen from the figure, mean AUC of the 10-fold cross-validation experiments in our study is 0.73, which means that there is a 73% chance that the model will be able to distinguish between consent and refusal utterances.

**Figure 6**

*AUC-ROC curve*

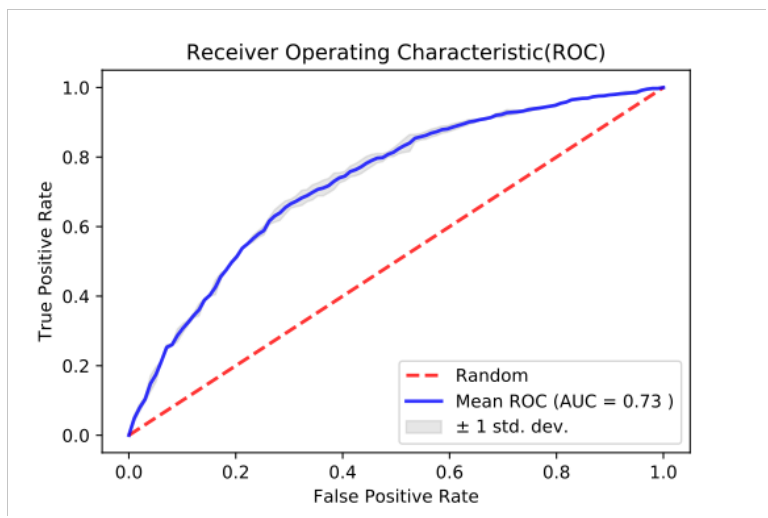

## 4. Discussion

### 4.1 Decreased Confidence in SA

We sought to analyze the acoustic signature of SA using two distinct, yet complementary approaches: a theory-driven (top-down) hypothesis testing approach and a data-driven (bottom-up) ML based approach. We hypothesized that, as compared to LSA, HSA individuals are associated with a confrontation-reduction signaling performance pattern. Interestingly, this was only partly supported by the theory-driven approach but more fully supported by the data-driven approach.

Specifically, using the GP classifier and taking the full breadth of vocal properties into account, we were able to distinguish between HSA and LSA with an accuracy of 75.6%, precision of 74% and recall of 69%. Further, our model evidenced good separability (.83). Moreover, using a decision tree technique, we concluded that HSA men were characterized by lower mean intensity, lower intensity variability, and higher shimmer than LSA men. Interestingly, HSA women were characterized by a different vocal pattern: a higher mF0, lower mean intensity and longer prompt-to-start as compared to LSA women. Our findings suggest that the acoustic signature of SA is gender-specific (Asher & Aderka, 2018), with *women (but not men)* exhibiting a vocal pattern consistent with low-dominance, conflict-avoidance strategies. In addition, consistent with our theoretical postulations, ML analysis also revealed that HSA participants were more likely to articulate differently than LSA participants when asked to express dominance (refuse a command) versus when needed to express appeasement (consent). Indeed, ML algorithms differentiated between HSA and LSA better (accuracy of 79%) using refusal, as compared to consent, utterances (accuracy of 70%).

This pattern of findings is consistent with the evolutionary model of SA, highlighting the enhanced use of conflict-reduction strategies in SA. According to the evolutionary model, HSA individuals evaluate themselves as low in social attractiveness, and fearing making bids for status or approval, as these claims may lead to loss of belongingness or exclusion. To remain a part of the group HSA individuals seek to de-escalate competition by using submissive or appeasing behaviors (Gilbert et al., 2001). In our case, these appeasing behaviors are reflected by lower mean intensity for both genders, as well as less dominant speech pattern, especially in women (great speech onset hesitance, higher pitch). By using ecologically valid, everyday tasks our findings support and extend previous findings (Galili et al., 2013; Salekin et al., 2018).

Consistent with our predictions and with the decision tree technique, statistical analysis showed that HSA participants demonstrated lower mean intensity than did LSA participants. However, inconsistent with our predictions and with previous research from our own, as well as other laboratories, our statistical analysis did not identify other low-dominant acoustic properties in HSA vs. LSA individuals. This inconsistency may be due to the specifics of our task and population. First, our task, which required participants to switch between consent and refusal, may have been cognitively taxing for all individuals, possibly decreasing individual differences in speech production. Second, whereas several previous studies used a task that enhanced state anxiety in socially anxious individuals by inviting them to perform a task that was highly stressful for them - public speaking (e.g., Salekin et al., 2018), we opted to engage our participants in less stressful, "ordinary" interpersonal task. Third, unlike previous studies, our sample consisted of non-clinical population, mostly women.

**4.2 Distinct Vocal Profiles of Refusal and Consent Utterances**

The second aim our study was to differentiate between refusal and consent utterances. Our hypothesized that consent is associated with a less confident pattern of vocal performance was supported by both top-down and bottom-up approaches. First, using top-down, classical hypothesis testing approach, we observed that consent utterances were characterized by higher F0 std, lower mean intensity, lower intensity std, higher jitter and higher shimmer, as compared to refusal utterances. These findings echo the patterns of acoustic profiles of commands and requests (Galili et al., 2013) as well as additional studies indicating that social dominance is associated with higher pitch variability (e.g. Leongoamez et al., 2017), lower speech intensity (e.g. Linnankoski et al., 2005), lower intensity variability (Tusing et al., 2000) and with lower jitter and shimmer (Shao et al., 2010). Second, using the GP classifier and taking the full breadth of acoustic properties into account, we were able to distinguish between refusal and consent utterances with an accuracy of 71.7%, precision of 72% and recall of 72%. Further, our model evidenced good separability (.73).

## 4.3 Clinical and Methodological Implications

Clinically, our research may contribute to the automatization of SAD diagnosis by identifying behavioral markers of SA-severity. The ability to diagnose SAD quickly and effectively is crucial, as 50% of individuals with this disorder never discuss their distress with a provider (Wang et al., 2005). The possibility of a diagnosis may encourage people to reach for professional help. Moreover, the results of the bottom-up analysis expand our understanding regarding the acoustic profiles of HSA individuals. These finding are in line with Gilbert's (2011) theory of impaired social status system in SAD. In our study, appeasement behaviors are reflected by lower mean intensity for both genders, as well as less dominant speech patterns, especially in women (more hesitance to start speaking, higher pitch). Importantly, lower mean intensity was also found using statistical analysis. Indeed, the findings of the current study add to the literature on gender differences in SA, suggesting that the vocal signature of SA is gender specific.

Methodologically, we utilized two practices of data analysis: theory-driven (top-down) hypothesis testing approach and data-driven (bottom-up) ML approach. The major difference between hypothesis-testing and ML is their purpose: statistical models are designed to enable *inference* about the relationships between variables, and so (relative) simplicity and good interpretability are basic features of such models. ML models are designed to make the most *accurate* predictions possible for new data based on known properties learned from the training data. Frequently, the interpretability of such models is weak. Combining the two approaches allowed us to accurately categorize individuals into two groups (HSA and LSA) while maintaining theory-based interpretability.

**4.4 Limitations and Future Directions**

Several limitations of our study should be noted. First, our studies were based on a non-clinical population and require a replication with a sample of clinically diagnosed patients. Second, we used a highly demanding task requiring participants to switch between two performance modes. This demanding task might have decreased group differences. Third, as we recorded the participants using a headset microphone, which position could shift, our intensity measures might have been affected by idiosyncratic changes due to movements towards or away from the microphone. Yet, we believe that a consistent bias that HSA individuals were positioned at a greater distance from the microphone in compared to LSA individuals is unlikely. Fourth, the aim of our study was to investigate the differences between HSA and LSA individuals. Therefore, we designed the study ensuring that we have an adequate number of participants in each group (HSA and LSA). However, deeper exploration of our data suggested that the acoustic signature of SA is gender specific. Given these findings, these results should be interpreted carefully due to the relatively small number of males participating in the study. In addition, given that HSA males may be related to dominance/submissiveness more than HSA females (e.g., see Maner, Miller, Schmidt & Eckel, 2008), the paucity of males' participants in our study may explain why our hypothesis were only partly supported using the top-down approach. Finally, complementing the analysis vocal properties with the concurrent analysis of verbal content may be needed. Indeed, as we already mentioned, longer speech duration increased dominance ratings for those who spoke with socially dominant language but decreased dominance ratings for those who did not use dominant-sounding language (Hodges-Simeon et al., 2010). Combining vocal and content analysis may improve our ability to differentiate HSA and LSA individuals.

**4.5 Summary and Conclusion**

In the current study we examined vocal patterns of everyday speech in HSA and LSA individuals using classical -- theory-driven, hypothesis-based -- approach as well as using a novel, data-driven approach. Consistent with the evolutionary model of SA, ML-based analytic approach (and, to a lesser degree, classical statistical techniques) documented that HSA, but not LSA individuals appears to be characterized by appeasing vocal patterns. Our best algorithm distinguished between HSA and LSA individuals with an accuracy of 75.6%. The present findings are promising with respect to the automatization of SAD diagnosis. The use of objective behavioral measures may significantly increase the number of people seeking help for SAD, as well as for other painful and debilitating conditions.

# References

Aderka, I. M., Hofmann, S. G., Nickerson, A., Hermesh, H., Gilboa-Schechtman, E. & Marom, S. (2012). Functional impairment in social anxiety disorder. *Journal of Anxiety Disorders*, *26*(3), 393-400.

Aderka, I. M., Weisman, O., Shahar, G. & Gilboa-Schechtman, E. (2009). The roles of the social rank and attachment systems in social anxiety. *Personality and Individual Differences, 47(4)*, 284–288.

Allan, S., & Gilbert, P. (1997). Submissive behaviour and psychopathology. *British Journal of Clinical Psychology, 36(4)*, 467-488.

American Psychiatric Association (2014). Diagnostic and Statistical Manual of Mental Disorders, Fifth Edition, Text Revision. Washington, DC: American Psychiatric Association.

Antony, M. M. (1997). Assessment and treatment of social phobia. *The Canadian Journal of Psychiatry, 42*(8), 826-834.

Antony, M. M., Coons, M. J., McCabe, R. E., Ashbaugh, A., & Swinson, R. P. (2006). Psychometric properties of the social phobia inventory: Further evaluation. Behaviour Research and Therapy, 44(8), 1177-1185.

Asher, M., & Aderka, I. M. (2018). Gender differences in social anxiety disorder. *Journal of clinical psychology, 74*(10), 1730-1741.

Baker, S. L., Heinrichs, N., Kim, H. J., & Hofmann, S. G. (2002). The Leibowitz social anxiety scale as a self-report instrument: A preliminary psychometric analysis. *Behaviour Research & Therapy, 40*, 701–715.

Banerjee, D., Islam, K., Xue, K., Mei, G., Xiao, L., Zhang, G., ... & Li, J. (2019). A deep transfer learning approach for improved post-traumatic stress disorder diagnosis. *Knowledge and Information Systems, 60*(3), 1693-1724.

Banse, R., & Scherer, K. (1996). Acoustic profiles in emotion expression. *Journal of Personality and Social Psychology, 70*, 614–636.

Boersma, P. & Weenik, D. (2009). Praat: doing phonetics by computer (Version 5.1. 03)[Computer program]. Retrieved March 21, 2009.

Bugental, D. B., Beaulieu, D. A., Schwartz, A., & Dragosits, R. (2009). Domain specific responses to power-based interaction. *Journal of Experimental Psychology, 45*, 386–391.

Cohn, J. F., Kruez, T. S., Matthews, I., Yang, Y., Nguyen, M. H., Padilla, M. T., ... & De la Torre, F. (2009, September). Detecting depression from facial actions and vocal prosody. In *2009 3rd International Conference on Affective Computing and Intelligent Interaction and Workshops* (pp. 1-7). IEEE.

Cowie, R., Douglas-Cowie, E., Tsapatsoulis, N., Votsis, G., Kollias, S., Fellenz, W., et al. (2001). Emotion recognition in human–computer interaction. *IEEE Signal Processing Magazine, 18(1)*, 32–80.

Cuming, S., Rapee, R. M., Kemp, N., Abbott, M. J., Peters, L., & Gaston, J. E. (2009). A self-report measure of subtle avoidance and safety behaviors relevant to social anxiety: Development and psychometric properties. *Journal of Anxiety Disorders, 23(7)*, 879-883.

Demetriou, C., Ozer, B. U., & Essau, C. A. (2014). Self-report questionnaires. *The encyclopedia of clinical psychology*, 1-6.

Edelmann, R. J., Asendorpf, J., Contarello, A., Georgas, J., Villanueva, C., & Zammuner, V. (1987). Self-reported verbal and non-verbal strategies for coping with embarrassment in five European cultures. *Information (International Social Science Council), 26(4)*, 869-883.

Erdogdu Sakar, B., Serbes, G., & Sakar, C. O. (2017). Analyzing the effectiveness of vocal features in early telediagnosis of Parkinson's disease. *PloS one*, 12(8), e0182428.

Erwin, B. A., Heimberg, R. G., Schneier, F. R., & Liebowitz, M. R. (2003). Anger experience and expression in social anxiety disorder: Pretreatment profile and predictors of attrition and response to cognitive-behavioral treatment. *Behavior Therapy, 34(3)*, 331-350.

Espinola, C. W., Gomes, J. C., Pereira, J. M. S., & dos Santos, W. P. (2020). Detection of major depressive disorder using vocal acoustic analysis and machine learning—an exploratory study. *Research on Biomedical Engineering*, 1-12.

Ferrand, C. T. (2001). Speech science: An integrated approach to theory and clinical practice. Ear and Hearing, 22(6), 549.

Fresco, D. M., Coles, M. E., Heimberg, R. G., Liebowitz, M. R., Hami, S., Stein, M. B., & Goetz, D. (2001). The Liebowitz social anxiety scale: A comparison of the psychometric properties of self-report and clinician-administered formats. *Psychological Medicine, 31*, 1025–1035.

Galili, L., Amir, O. & Gilboa-Schechtman, E. (2013). In a high pitched voice: Acoustic properties of dominance utterances in social anxiety. *Journal of Social and Clinical Psychology, 32*, 651-673

Gilbert, P. (2001). Evolution and social anxiety: The role of attraction, social competition, and social hierarchies. *Psychiatric Clinics of North America, 24(4)*, 723–751.

Gorman, J. M., & Gorman, L. K. (1987). Drug treatment of social phobia. *Journal of Affective Disorders, 13*(2), 183-192.

Hall, J. A., Coats, E. J., & LeBeau, L. S. (2005). Nonverbal behavior and the vertical dimension of social relations: a meta-analysis. *Psychological bulletin, 131(6)*, 898.

Haghani, P., Narayanan, A., Bacchiani, M., Chuang, G., Gaur, N., Moreno, P., ... & Waters, A. (2018, December). From audio to semantics: Approaches to end-to-end spoken language understanding. In *2018 IEEE Spoken Language Technology Workshop (SLT)* (pp. 720-726). IEEE.

Hodges-Simeon, C. R., Gaulin, S. J., & Puts, D. A. (2010). Different vocal parameters predict perceptions of dominance and attractiveness. *Human Nature, 21(4)*, 406-427.

Maner, J. K., Miller, S. L., Schmidt, N. B., & Eckel, L. A. (2008). Submitting to defeat: Social anxiety, dominance threat, and decrements in testosterone. *Psychological Science, 19*(8), 764-768.

Johnson, S. L., Leedom, L. J. & Muhtadie, L. (2012). The dominance behavioral system and psychopathology: Evidence from self-report, observational, and biological studies. *Psychological Bulletin*, *138(4)*, 692–743.

Jones, B. C., Feinberg, D. R., DeBruine, L. M., Little, A. C., & Vukovic, J. (2010). A domain-specific opposite-sex bias in human preferences for manipulated voice pitch. *Animal Behaviour, 79*, 57–62.

Kimble, C. E., & Musgrove, J. I. (1988). Dominance in arguing mixed-sex dyads: Visual dominance patterns, talking time, and speech loudness. *Journal of Research in Personality, 22*(1), 1-16.

König, A., Satt, A., Sorin, A., Hoory, R., Toledo-Ronen, O., Derreumaux, A., ... & David, R. (2015). Automatic speech analysis for the assessment of patients with predementia and Alzheimer's disease. *Alzheimer's & Dementia: Diagnosis, Assessment & Disease Monitoring, 1*(1), 112-124.

Laukka, P., Linnman, C., Åhs, F., Pissiota, A., Frans, Ö., Faria, V., et al. (2008). In a nervous voice: Acoustic analysis and perception of anxiety in social phobics' speech. *Journal of Nonverbal Behavior, 32*, 195–214.

Lazarov, A., Marom, S., Yahalom, N., Pine, D. S., Hermesh, H., & Bar-Haim, Y. (2018). Attention bias modification augments cognitive–behavioral group therapy for social anxiety disorder: a randomized controlled trial. *Psychological medicine*, *48*(13), 2177.

Leinonen, L., Hiltunen, T., Linnankoski, I., & Laakso, M. L. (1996). Expression of emotional-motivational connotations with a one-word utterance. *Journal of the Acoustical Society of America, 102*, 1853–1863.

Leongómez, J. D., Mileva, V. R., Little, A. C., & Roberts, S. C. (2017). Perceived differences in social status between speaker and listener affect the speaker's vocal characteristics. *PloS one, 12(6)*, e0179407.

Lewin, M. R., McNeil, D. W., & Lipson, J. M. (1996). Enduring without avoiding: Pauses and verbal dysfluencies in public speaking anxiety. *Journal of Psychopathology & Behavioral Assessment, 18*, 387–402.

Liebowitz, M. R. (1987). Social phobia. *Modern Problem of Pharmacopsychiatry, 22*, 141–173.

Linnankoski, I., Leinonen, L., Vihla, M., Laakso, M. L., & Carlson, S. (2005). Conveyance of emotional connotations by a single word in English. *Speech Communication, 45*, 27–39.

Maxhuni, A., Muñoz-Meléndez, A., Osmani, V., Perez, H., Mayora, O., & Morales, E. F. (2016). Classification of bipolar disorder episodes based on analysis of voice and motor activity of patients. *Pervasive and Mobile Computing, 31*, 50-66.

Mennin, D. S., Fresco, D. M., Heimberg, R. G., Schneier, F. R., Davies, S. O., & Liebowitz, M. R. (2002). Screening for social anxiety disorder in the clinical setting: using the Liebowitz Social Anxiety Scale. *Journal of anxiety disorders*, *16*(6), 661-673.

Oakman, J., Gifford, S., & Chlebowsky, N. (2003). A multilevel analysis of the interpersonal behavior of socially anxious people. *Journal of Personality, 71(3)*, 397-434.

Ohala, J. J. (1984). An ethological perspective on common cross—language utilization of F0 of voice. *Phonetica, 41*, 1–16.

Petersen, G. E., & Barney, H. L. (1952). Control methods used in a study of the vowels. *Journal of the Acoustical Society of America, 24*, 175–184.

Price, R., Mehrabani, M., & Bangalore, S. (2020, May). Improved end-to-end spoken utterance classification with a self-attention acoustic classifier. In *ICASSP 2020-2020 IEEE International Conference on Acoustics, Speech and Signal Processing (ICASSP)* (pp. 8504-8508). IEEE.

Puts, D. A., Hodges, C. R., Cárdenas, R. A., & Gaulin, S. J. (2007). Men's voices as dominance signals: vocal fundamental and formant frequencies influence dominance attributions among men. Evolution and Human Behavior, 28(5), 340-344.Puts, D. A., Gaulin, S.J.C., & Verdolini, K. (2006) Dominance and the evolution of sexual dimorphism in human voice pitch. *Evolution and Human Behavior, 27*, 283–96.

Rochman, D., & Amir, O. (2013). Examining in-session expressions of emotions with speech/vocal acoustic measures: An introductory guide. *Psychotherapy Research*, *23*(4), 381-393.

Ruscio, A. M., Brown, T. A., Chiu, W. T., Sareen, J., Stein, M. B. & Kessler, R. C. (2008). Social fears and social phobia in the USA: Results from the National Comorbidity Survey Replication. *Psychological Medicine, 38,* 5-28.

Salekin, A., Eberle, J. W., Glenn, J. J., Teachman, B. A., & Stankovic, J. A. (2018). A weakly supervised learning framework for detecting social anxiety and depression. *Proceedings of the ACM on interactive, mobile, wearable and ubiquitous technologies, 2*(2), 1-26.

Serdyuk, D., Wang, Y., Fuegen, C., Kumar, A., Liu, B., & Bengio, Y. (2018, April). Towards end-to-end spoken language understanding. *IEEE International Conference on Acoustics, Speech and Signal Processing (ICASSP)* (pp. 5754-5758).

Schneier, F. R., Kent, J. M., Star, A. & Hirsch, J. (2009). Neural circuitry of submissive behavior in social anxiety disorder: a preliminary study of response to direct eye gaze. *Psychiatry Research: Neuroimaging, 173(3)*, 248-250.

Shao, J., MacCallum, J. K., Zhang, Y., Sprecher, A., & Jiang, J. J. (2010). Acoustic analysis of the tremulous voice: assessing the utility of the correlation dimension and perturbation parameters. *Journal of communication disorders*, *43*(1), 35-44.

Silber-Varod, V., Kreiner, H., Lovett, R., Levi-Belz, Y., & Amir, N. (2016). Do social anxiety individuals hesitate more? The prosodic profile of hesitation disfluencies in Social Anxiety Disorder individuals. *Proceedings of Speech Prosody (SP2016)*, 1211-1215.

Sobol-Shikler, T. (2009). Analysis of affective expression in speech (No. UCAM-CL-TR-740). University of Cambridge, Computer Laboratory.

Trainor, L. J., Austin, C. M., & Desjardins, R. N. (2000). Is infant-directed speech prosody a result of the vocal expression of emotion? *Psychological science, 11(3)*, 188-195.

Tusing, K. J., & Dillard, J. P. (2000). The sounds of dominance: Vocal precursors of perceived dominance during interpersonal influence. *Human Communication Research, 26*, 148–171.

Vergyri, D., Knoth, B., Shriberg, E., Mitra, V., McLaren, M., Ferrer, L., ... & Marmar, C. (2015). Speech-based assessment of PTSD in a military population using diverse feature classes. In *Sixteenth annual conference of the international speech communication association*.

Weeks, J. W., Lee, C.-Y., Reilly, A. R., Howell, A. N., France, C., Kowalsky, J. M. & Bush, A. (2012). "The Sound of Fear": Assessing vocal fundamental frequency as a physiological indicator of social anxiety disorder. *Journal of Anxiety Disorders, 26(8)*, 811–822.

Weeks, J. W., Heimberg, R. G. & Heuer, R. (2011). Exploring the role of behavioral submissiveness in social anxiety. *Journal of Social and Clinical Psychology*, *30(3)*, 217–249.

Wang, P.S., Berglund, P., Olfson, M., Pincus, H.A., Wells., K.B., & Kessler, R.C. (2005). Failure and delay in initial treatment contact after first onset of mental disorders in the National Comorbidity Survey Replication. *Archives of General Psychiatry 62*(6), 603–613.

Wong, N., Sarver, D. E. & Beidel, D. C. (2012). Quality of life impairments among adults with social phobia: The impact of subtype. *Journal of Anxiety Disorders*, *26(1)*, 50-57.

Zuroff, D. C., Fournier, M. A., Patall, E. A. & Leybman, M. J. (2010). Steps toward an evolutionary personality psychology: Individual differences in the social rank domain. *Canadian Psychology/Psychologie Canadienne, 51(1)*, 58-66.

Acknowledgements

1. This work was supported by the Israel Science Foundation Grant number 740/15 awarded to Eva Gilboa-Schechtman. Sources of financial support had no influence over the design, analysis, interpretation, or choice of submission outlet for this research.

2. Data are available by contacting the corresponding author.